\documentstyle[12pt]{article}

\def\boxit#1{
\vbox{\hrule height0.5pt\hbox{\vrule width0.5pt\kern10pt\vbox{
\kern10pt#1\kern10pt}
\kern10pt\vrule width0.5pt}\hrule height0.5pt}}

\def\bild#1\over#2{\mathrel{\mathop{\kern5pt #1}\limits_{#2}}}

\newcommand{\be}{\begin{equation}}
\newcommand{\ee}{\end{equation}}
\newcommand{\bc}{\begin{center}}
\newcommand{\ec}{\end{center}}
\newcommand{\ba}{\begin{array}}
\newcommand{\ea}{\end{array}}

\begin{document}

\title{ Self Similar Solutions of the Evolution Equation\\
of a Scalar Field in an Expanding Geometry}
\author{F. Braghin\thanks{Doctoral fellow of Coordena\c c\~ao de
 Aperfei\c coamento de Pessoal de N\'\i vel Superior,Brasil }
, C. Martin and D. Vautherin \\
Division de Physique Th\'eorique\thanks{Unit\'e de Recherche des
Universit\'es Paris XI et Paris  VI associ\'ee au CNRS},\\
 Institut de Physique Nucl\'eaire, 91406, Orsay Cedex, France
}
\date{}
\maketitle

{\bf Abstract}:   We consider the functional Schr\"odinger equation for a
self interacting scalar field in an expanding geometry. By performing a time
dependent scale transformation on the argument of the field we derive a
functional Schr\"odinger equation whose hamiltonian is time independent but
involves a time-odd term associated to a constraint on the expansion current.
We study the mean field approximation to this equation and generalize in this
case, for interacting fields, the solutions worked out by Bunch and Davies for
free fields.

\vspace*{5cm}

\noindent IPNO/TH 94-39\hfill{May 1994}

\newpage
  Several authors have pointed out that inflationary models based on effective
potentials ignore the fact that the very concept of effective potential implies
a static approximation which may be invalidated by dynamical effects
\cite{DEVEGA}. The purpose of this present letter is to investigate this
question using  the functional Schr\"odinger equation for a
scalar field in an expanding geometry:
\be \label{1}
i \hbar \partial_t \Phi [\varphi]= \frac{1}{2} \int d {\bf x}
\{- \frac{\hbar^2}{a^3(t)} \frac{\delta^2 \Phi}{\delta \varphi({\bf x})
\delta \varphi({\bf x})}+[a(t)( \nabla \varphi({\bf x}))^2+
a^3(t)(m_0^2 \varphi^2({\bf x})+\frac{\lambda}{24} \varphi^4({\bf x}))] \Phi
\}.
\ee
In this equation $a(t)=\exp(\chi t)$ specifies the metric $ds^2=
dt^2-a^2(t) d{\bf x}^2$ and $\chi$ is the expansion parameter
(Hubble's constant). The constants $m_0$ and $\lambda$ are respectively the
bare mass of the field and the bare coupling constant. The evolution
equation (\ref{1}) was considered by Guth and Pi \cite{GUTH} and by
Eboli, Jackiw and So- Young Pi \cite{JACKIW}. In particular in reference
\cite{JACKIW} a numerical solution of this equation was attempted in the
mean field approximation. However instabilities were found in the
implementation
of the method. Our aim in the present study is to perform a particular scale
transformation on the wave functional which brings equation (\ref{1}) into
the form of a static functional Schr\"odinger equation. The method, although
more elaborate, is reminiscent of the transformation used in nuclear
physics to describe rotating nuclei \cite{THOULESS}.

Let us look for solutions of equation (\ref{1}) in the form
\be \label{2}
\Phi [\varphi({\bf x}),t] = \Psi [ \xi_{\alpha}({\bf x}),t] \frac{1}{\sqrt
{N(\alpha)}},
\ee
where
\be \label{3}
\xi({\bf x})= \varphi(\alpha(t) \: {\bf x}),
\ee
and where $N(\alpha)$ is a
normalisation factor. By using the definition of the functional derivative
we first find
\be \label{4}
\delta \Phi= \int \frac{\delta \Psi}{\delta \xi} \delta \varphi(\alpha {\bf x})
d{\bf x}=  \int \frac{\delta \Psi}{\delta \xi({\bf y}/\alpha)}
\delta \varphi({\bf y}) \alpha^{-3} d{\bf y},
\ee
which allows one to express the functional derivative of $\Phi$ in terms of the
functional derivative of $\Psi$ as
\be \label{5}
\frac{\delta \Phi}{\delta \varphi({\bf x})}= \frac{1}{\alpha^3}
\frac{\delta \Psi}{\delta \xi({\bf x}/\alpha)}.
\ee
By applying this formula twice we find that the kinetic energy of the field
is
\be \label{6}
\int d{\bf x} \frac{\delta^2 \Phi}{ \delta \varphi({\bf x})
\delta \varphi({\bf x}) }= \frac{1}{\alpha^3}
\int d{\bf x} \frac{\delta^2 \Psi}{ \delta \xi({\bf x})
\delta \xi({\bf x}) }.
\ee
The transformation law for the mass term is easily derived from the relation
\be \label{7}
\int d{\bf x} \: \varphi^2({\bf x})=
\alpha^3 \int d{\bf y} \: \varphi^2(\alpha {\bf y})=
\alpha^3 \int d{\bf y} \: \xi^2({\bf y}),
\ee
while for the gradient term we have
\be \label{8}
\int d{\bf x} \: (\nabla \varphi ({\bf x}))^2=
\frac{\alpha^3}{\alpha^2} \int d{\bf y} \: (\nabla_{\bf y}
\varphi(\alpha {\bf y}))^2=
\alpha \int d{\bf y} \: (\nabla \xi({\bf y}))^2.
\ee
We still have to calculate the time derivative of the wave funtional $\Phi$.
Between $t$ and $t+dt$ the change $\delta \Phi$ in $\Phi$ is obtained by
combining the time change of $\Psi$ and the change in $\alpha(t)$ with the
result

\be \label{9}
\delta \Phi= \frac{\partial \Psi}{\partial t} dt+
\int \frac{\delta \Psi}{\delta \xi({\bf x})} \delta \xi({\bf x})
d{\bf x} -\frac{\delta N}{2 N} \Psi \ .
\ee
Since the change in $\xi$ is
\be \label{10}
\delta \xi({\bf x})=\varphi(\alpha{\bf x}+ {\bf x} \delta \alpha)
-\varphi(\alpha {\bf x}) \\
=\xi({\bf x}+ \frac{\delta \alpha}{\alpha} {\bf x})- \xi({\bf x})\\
=\frac{\delta \alpha}{\alpha} {\bf x}.\nabla \xi \ ,
\ee
we obtain
\be \label{11}
\partial_t \Phi= \partial_t \Psi + i\frac{\dot \alpha}{\alpha} \int d{\bf x}
\pi({\bf x}) {\bf x}.\nabla \xi \Psi- \frac{\dot N}{2N} \Psi \ .
\ee
The time derivative of $N(\alpha)$ is easily calculated in a similar way.
The final
result is
\be \label{12}
\partial_t \Phi= \partial_t \Psi + \frac{i}{2}
\frac{\dot \alpha}{\alpha} \int d{\bf x}
\{ \pi({\bf x}) {\bf x}.\nabla \xi + {\bf x}.\nabla \xi \pi({\bf x}) \} \Psi
\ .
\ee
By choosing $\alpha(t)=1/a(t)=\exp(-\chi t)$,
we thus see that $\Phi$ is a solution of the
time-dependent Schr\"odinger equation (\ref{1}) provided $\Psi$ satisfies
\be \label{13}
i \partial_t \Psi [\xi] = H_{\chi} \Psi [\xi],
\ee
where $H_{\chi}$ is the time- independent hamiltonian
\be \label{14}
H_{\chi}= \int d{\bf x} \{- \frac{1}{2} \frac{\delta^2}{
\delta \xi({\bf x}) \delta \xi({\bf x}) }
+\frac{(\nabla \xi)^2}{2}+ m_0^2 \xi^2+ \frac{\lambda}{24} \xi^4
+\frac{i \chi}{2}
\left[ \pi({\bf x}) ({\bf x}.\nabla \xi) + ({\bf x}.\nabla \xi) \pi({\bf x})
\right] \} .
\ee
The last term in the previous equation is static. However it corresponds
to a constraint on the expectation value of the expansion current.
 Indeed in ordinary quantum mechanics of a single particle
this operator would be similar to the operator $ {\bf x . p} + {\bf p . x}$.

The presence of this term illustrates the relevance of the discussion
given in reference \cite{DEVEGA} about the importance of dynamical effects.
Indeed, as will be shown below, this term affects the gap equation and may thus
drive the system out of the symmetric or asymmetric phase.

In order to obtain a mean field approximation to the functional Schrodinger
 equation (13) we assume that at each time t the wave functional describing
the system is a gaussian functional of the form:
\be \label{15}
\Psi [ \xi ,t ] = \ba{ll} & {\displaystyle
{\cal  N}
\exp \left\{ i \int d{\bf x} \: \bar \pi ({\bf x},t) \:
\left(\xi({\bf x})-\bar \xi({\bf x},t) \right) \right\}
} \\
& {\displaystyle \times
\exp \left\{ - \int d{\bf x} \: d{\bf y} \: \left( \xi({\bf x}) -
\bar \xi({\bf x},t) \right) \: \left(\frac{1}{4} G^{-1}({\bf x},{\bf y}) + i
\Sigma({\bf x},{\bf y},t) \right) \: \left( \xi({\bf y})-\bar \xi({\bf y},t)
\right) \right\}
 } \ , \ea
\ee
where ${\cal N}$ is a normalization factor.
Our trial functional contains four variational parameters:
 the center of the gaussian $ \bar \xi({\bf x},t) $,
its conjugate momentum $ \bar \pi({\bf x},t) $, and the
 real and imaginary parts of its kernel: $ G({\bf x,y},t)$ and $ \Sigma ({\bf
x,y},t) $.

The expectation value of the hamiltonian $ H_{\chi }$ evaluated with the
gaussian trial state $ \Psi $ is
\be \label{16}
\ba{ll}
<\Psi | H_{\chi } |\Psi > = &\int \: d{\bf x} \left\{ \frac{1}{2} \bar \pi^2 +
 \frac{1}{2}(\nabla \bar \xi)^2 + \frac{1}{2} m_0^2 \bar \xi^2 +
\frac{1}{24} \lambda \bar \xi^4 \right.
\\
& \left. +  \frac{1}{2} \left[ \frac{1}{4} G^{-1}({\bf x},{\bf x},t)
+ 4 \left(G \Sigma G\right) ({\bf x},{\bf x},t) +
\left(-\Delta +m_0^2+ \frac{1}{2} \bar \xi^2+
\frac{1}{4} \lambda G({\bf x},{\bf x},t )\right) G({\bf x},{\bf x},t) \right]
\right. \\
& \left. + \frac{\chi}{2} \left[
 \: \bar \pi({\bf x},t) \: {\bf x} . \nabla \bar \xi({\bf x,t}) -
\bar \xi({\bf x},t) \: \nabla . {\bf x} \: \bar \pi({\bf x},t) \right] \right.
\left. - \chi \:
{\rm Trace} \{ \Sigma ({\bf x}. \nabla) G- G ( \nabla.{\bf x}) \Sigma \}
\right\}  \ .
\ea
\ee

By varying the action \cite{JACKKERMAN}
$<\Psi | i\partial_{t} - H_{\chi } |\Psi >$
with respect
to the variational parameters, we obtain the following dynamical equations :
\be  \label{17}
\ba{ll}
\partial_t G=& 2 (G \Sigma +\Sigma G) -
\chi \{ ({\bf x}. \nabla) G- G (\nabla.{\bf x}) \},
\\
\partial_t \Sigma=& \frac{1}{8} G^{-2} - 2\Sigma^2 -
\frac{1}{2} \{ -\Delta + m_0^2 + \frac{1}{2} \lambda \bar \xi^2 +
\frac{1}{2} \lambda \: G({\bf x},{\bf x}) \}
\\
& - \chi \left\{ (\Sigma {\bf x} . \nabla) -\Sigma (  \nabla.{\bf x}) \right\}
\\
\partial_t {\bar \xi}=& -\bar \pi + \chi ({\bf x} . \nabla \bar \xi),
\\
\partial_t {\bar \pi}=& \chi ({\bf x} . \nabla \bar \pi) +
\{ -\Delta + m_0^2 + \frac{1}{6} \lambda \bar \xi^2 +
\frac{1}{2} \lambda \: G({\bf x},{\bf x}) \} \bar \xi .
\ea
\ee
To derive these expressions  one has to remember that the cyclic
invariance of the trace does not hold for pairs of operators such as {\bf x}
and {\bf p}.

The previous equations admit translationally invariant solutions of the form
\be \label{17a}
\ba{ll}
\bar \xi({\bf x},t) =& \bar \xi(t) \ ,
\\
\bar \pi({\bf x},t) =&  \bar \pi(t) \ ,
\\
G({\bf x,y}) =& \frac{1}{(2 \pi)^3} \int d{\bf k} \:
G(k) \exp(i{\bf k}({\bf x-y})) \ ,
\\
\Sigma({\bf x,y}) =& \frac{1}{(2 \pi)^3} \int d{\bf k} \: \Sigma(k)
\exp(i{\bf k}({\bf x-y})) \ ,
\ea
\ee
provided that $\bar \xi$, $\bar \pi$, $G(k)$ and $\Sigma(k)$ satisfy :
\be \label{18}
\ba{ll}
\partial_t G(k) =& 4 G(k)\Sigma(k) -
\chi \{ ( {\bf k} . \nabla_k) G(k) + 3 G(k) \} \ ,
\\
\partial_t\Sigma(k) =& \frac{1}{8} G^{-2} (k) - 2 \Sigma^2 (k) +
\frac{1}{2}
\{ {\bf k}^2 + m_0^2 + \frac{1}{2} \lambda \bar \xi^2 +
\frac{\lambda}{2} \frac{1}{(2\pi)^3} \int d{\bf k} G(k) \} -
\chi \{  3 \Sigma(k) - {\bf k} . \nabla_k \Sigma(k) \} \ ,
\\
\partial_t {\bar \xi}=&- \bar \pi  ,
\\
\partial_t {\bar \pi}=&
\left( m_0^2 + \frac{1}{6} \lambda \bar \xi^2 +
\frac{\lambda}{2} \frac{1}{(2\pi)^3} \int d{\bf k} \: G(k)\right) \: \bar \xi
 \ .
\ea \ee
Static solutions in the expanding frame correspond, after elimination of
$\bar \pi$ and $\Sigma$ to :
\be \label{19}
\left\{ m_0^2 + \frac{\lambda}{6}\bar \xi^2 + \frac{\lambda}{2}
\frac{1}{(2 \pi)^3}
\int d^3k \: G(k) \right\} \bar \xi =0
\ee
and
\be \label{20}
G^{-1} \: k^2 \frac{d^2}{d k^2} G + G^{-1} \: k \frac{d}{dk} G -
\frac{1}{2} \: G^{-2} \: k^2 \left( \frac{d}{dk} G \right)^2
-\frac{1}{2 \chi^2 G^2} + 2 \left( \frac{k^2}{\chi^2}-\nu^2 \right) = 0
\ , \ee
where $\nu$ satisfies the self-consistent equation :
\be \label{21}
\nu^2 \chi^2 = \frac{9}{4}\chi^2 - m_0^2 - \frac{\lambda}{2}\bar \xi^2 -
 \frac{\lambda}{2} \frac{1}{(2\pi)^3}\int d{\bf k} \: G({\bf k}) \ .
\ee
Equation (\ref{20}) is the differential equation  for the square root of the
modulus of the Hankel Function, and the most general solution has the form:
\be \label{22}
G(k) = \frac{\pi}{4\chi}\left| c_1 H^{(1)}_{\nu}(k / \chi) +
 c_2 H^{(2)}_{\nu}(k/ \chi )\right|^2
\ee
where $H^{(i)}$ are the first (i=1) and second (i=2) Hankel function of
order $\nu$, and the coeficients $c_1$ and $c_2$ obey the relation:
$|c_1|^2 - |c_2|^2 = 1$, and are determined by the boundary conditions.
 If we impose that the asymptotic behavior be the same as in flat space-time
 we get $c_1 = 1$ and $c_2 = 0$ \cite{GUTH}.

Equation (\ref{21})
will be refered to as the gap equation since it generalizes the
equation obtained for the mass gap of a self interacting scalar field in mean
field theory in Minkowski space. This limiting case is obtained for
$\chi = 0$.  In the limit $\chi=0$, $\nu$ is large and imaginary and
behaves as $i m / \chi$ where $m$ is the mass gap at $\chi=0$. Using
the asymptotic formula \cite{ABRAMO}
\be \label{22a}
\left|  H^{(1)}_{i \nu}( \nu / {\rm sinh} \alpha) \right|^2 \simeq
\frac{1}{ \pi \nu {\rm tanh } \alpha} \ ,
\ee
one obtains $G(k)=1/2 \sqrt{k^2+ m^2}$ as expected.

Inserting the gap equation into the equation for the background field
$\bar \xi$ we find that static solutions are determined by the two
conditions :
\be \label{23}
\nu^2\chi^2 -\frac{9}{4}\chi^2+ m^2_0 + \frac{\lambda}{2}\bar \xi^2 +
\frac{\pi}{8\chi} \lambda  \frac{1}{(2 \pi)^3}
\int d{\bf k}\left|H^{(1)}_{\nu}\right|^2 = 0
\ee
and
\be \label{24}
(\nu^2\chi^2 + \frac{\lambda}{3}\bar \xi^2-\frac{9}{4}\chi^2)\:
\bar \xi = 0 \ .
\ee
Before discussing the properties of these two equations it is worthwhile
examining the structure of the associated solutions when they are rewritten
in the laboratory frame. Denoting by $\tilde G$ the kernel of the
gaussian in this frame we have
$\tilde G(k)=e^{-3\chi t} G(k e^{-\chi t}) $, i. e.
\be \label{25}
\tilde G(k,t)=
e^{-3\chi t} \frac{\pi}{4\chi} \left|H^1_{\nu}(ke^{-\chi t})\right|^2
\ . \ee
For a free scalar field, $\nu$ is given by
\be \label{26}
\nu^2 \chi^2 = \frac{9}{4} \chi^2 -m_0^2
\ee
and the expression (\ref{25}) is identical to    the solution obtained
 by Bunch and Davies \cite{BUNCH}. In our formalism
this solution appears as a static solution
 in the expanding background geometry. However, our solution (\ref{25})
holds also for the case of a self interacting scalar field.
Different classes of solutions with broken or unbroken symmetry can be found.

As for the case $\chi=0$ the gap equation contains divergent quantities
and must be renormalized. In order to investigate this question let us
introduce a momentum cutoff $\Lambda$. At large momentum, using the asymptotic
form of Hankel functions, one finds that $G(k)$ behaves as
\be \label{26}
G(k) \simeq \frac{1}{2k} \left( 1- \frac{\chi^2 - 4 \nu^2 \chi^2}{8k^2}
\right),
\ee
so that the quantity $G$ contains a quadratic and a logarithmic divergence
when $\Lambda$ goes to infinity
\be \label{27}
G(\nu^2 \chi^2)= \frac{1}{8 \pi^2} \left\{ \Lambda^2
-\frac{\chi^2 -4 \nu^2 \chi^2}{8}
\log (\frac{16 \Lambda^2}{e (\chi^2- 4 \nu^2 \chi^2) } ) \right\}.
\ee
The key to the renormalization of our variational calculation is the existence,
for a given value of $\lambda$, of a critical value $r_c$ of the ratio $r_0=
m_0^2/ \Lambda^2$ beyond which there is no asymmetric solution of the gap
equation \cite{KERMAN}. For a large value of the momentum cutoff,
the gap equation can indeed be written as
\be \label{28}
- \nu^2 \chi^2+ \frac{\chi^2}{4}= \epsilon \mu^2 + \lambda_R \bar \xi^2+
\frac{\lambda_R}{32 \pi^2} (- \nu^2 \chi^2+ \frac{\chi^2}{4})
\log \left( \frac{\chi^2- 4 \nu^2 \chi^2}{\mu^2} \right),
\ee
where $\epsilon= \pm 1$ specifies wether $r_0$ is above or below $r_c$
and where we have introduced the following quantities
\be \label{29}
\lambda_R= \frac{\lambda}{
2+\frac{\lambda}{16 \pi^2} \log \left( \frac{4 \Lambda^2}{e \mu^2}  \right)
},
\ee
and
\be \label{30}
\epsilon \frac{\mu^2}{\Lambda^2}= \frac{2 \lambda_R}{\lambda}
(r_0+ \frac{\lambda}{16 \pi^2}- \frac{2 \chi^2}{ \Lambda^2} ),
\ee
i.e.
\be \label{31}
\epsilon \frac{\mu^2}{\Lambda^2}= \frac{2 \lambda_R}{\lambda}
(r_0 - r_c).
\ee
Keeping the ultraviolet cutoff $\Lambda$ fixed, the approach to the
critical point depends on the value of the expansion parameter $\chi$.

Below $r_c$ (i.e. $\epsilon=- 1$) there is a solution with broken
symmetry while for $r_0>r_c$ there is only a symmetric solution \cite{KERMAN}.
For a given value of $\lambda$,
approaching
the critical point from the left the solution of the gap equation is
\be \label{32}
-\nu^2 \chi^2 + \frac{\chi^2}{4}= \frac{4}{e} \:
\Lambda^2 e^{-\frac{16 \pi^2}{\lambda} }.
\ee
We thus see that when $\Lambda$ goes to infinity this solution remains finite
provided that the bare coupling constant $\lambda$ evolves as
\be \label{33}
\frac{1}{\lambda}= \frac{1}{16 \pi^2} \log
\left( \frac{\Lambda^2}{M_R^2} \right),
\ee
where $M_R$ is an arbitrary scale.

In conclusion, we have been able to find self similar solutions of the
mean-field evolution equation for the fluctuations of a self interacting scalar
field in an expanding geometry. This has been possible in the particular case
of a de Sitter geometry for which the expansion rate $\dot a /a$ is a
constant. In the case of a radiation dominated univers for which
$a(t)=\sqrt t$, this type of solutions can not longer be found. However,
in this case one has $\dot a/a=1/2 t$ and
the formalism we have described can be used to
make  predictions about the long time behaviour.

{\bf Acknowledgements} We are grateful to Alain Comtet for stimulating
discussions.

\end{document}